
\input amstex
\documentstyle{amsppt}
\overfullrule=0pt
\TagsOnLeft
\CenteredTagsOnSplits

\def\C{{\Bbb C}}
\def\Z{{\Bbb Z}}
\def\P{{\Bbb P}}
\def\O{{\Cal O}}
\def\Oh{{\widehat{\O}}}

\def\h{{\frak h}}
\def\g{{\frak g}}
\def\gh{{\widehat{\frak g}}}

\def\s{s}

\def\t{t}

\def\l{\ell}
\def\tV{\tilde{V_\vlm}}
\def\W{{\Cal W}}
\def\U{{\Cal N}}
\def\H{{\Cal H}}
\def\Hl{{\Cal H_{\lm}}}
\def\Hld{{\Cal H_{\lm}^\dagger}}
\def\Hv{\H_{\vec \lm}}
\def\F{{\Cal E}}
\def\fF{{\Cal F}}
\def\gr{\hbox{gr}^{\F}}
\def\Gr{\hbox{gr}^{\fF}}

\def\vlm{{\vec \lm}}
\def\Vac{{\Cal V}}

\def\X{{\frak X}}

\def\-{.}
\def\th{\theta}
\def\lm{\lambda}
\def\al{\alpha}
\def\Cx{\C ((\xi))}
\def\Cj{\C ((\xi_j))}
\def\<{\langle}
\def\>{\rangle}

\def\Xj{X\otimes\xi_j}
\def\j{\otimes\xi_j}

\def\Hom{\hbox{\rm Hom}\,}
\def\Res{\hbox{\rm Res}\,}
\rightheadtext{conformal blocks in the WZNW-models}
\topmatter
\title
 Finite-Dimensionality of
 the space of Conformal Blocks
\endtitle
\author  Takeshi Suzuki  \endauthor
\affil
Research Institute for Mathematical Sciences,
Kyoto University,
Kyoto, 606-01,
Japan
\endaffil

\abstract
Without using Gabber's theorem, the finite-dimensionality of the space of
conformal blocks in
the Wess-Zumino-Novikov-Witten models
is proved.
\endabstract

\endtopmatter

\document

\subhead
\S0 Introduction
\endsubhead
\medskip
Conformal field theory
with non-abelian gauge symmetry, called
the Wess-Zumino-Novikov-Witten (WZNW) model, has been studied
by many physicists and mathematicians.
A mathematical formulation of this model over the projective line is given
in [TK], and
it is generalized in [TUY] over algebraic curves of arbitrary genus.
In (chiral) conformal field theory, the main objects are $N$-point functions,
 and
in [TUY] they are regarded as sections
of a certain vector bundle over the moduli space of $N$-pointed stable curves.
The fiber of this bundle at a stable curve $\X$ is called
 the space of conformal blocks (or the space of vacua) attached to $\X$.
The finite-dimensionality of this space
is essential to the mathematical treatment of conformal field theory,
and is proved in [TUY] as a consequence of Gabber's theorem [Ga]
stating the involutivity of characteristic varieties.

The aim of the present paper is to
give a proof of the finite-dimensionality
without using Gabber's theorem.

In \S1 we summarize some basic facts on affine Lie algebras following [Ka].
In \S2 and \S3, we recall the definition of pointed stable curves and
the space of conformal blocks.
The essence in our proof is explained in \S 4
and the complete proof is given in \S 5.

\subhead
\S1. Integrable highest weight modules of Affine Lie algebras
\endsubhead
\medskip
By $\C [[\t]]$ and $\C ((\t))$, we mean the ring of formal power series
in $\t$ and the field of formal Laurent series in $\t$, respectively.
Let $\g$ be a simple Lie algebra over $\C$,
$\h$ its Cartan subalgebra and $\h^*$ the dual space of $\h$ over $\C$.
By
$\triangle$, we denote the root system of $(\g,\h)$, and
for $\al\in\triangle$ we denote
the root vector corresponding to $\al$ by $X_\al$.
Let
$$
(\ ,\ ) : \g \times \g \to \C
$$
be the Cartan-Killing form normalised by the condition
$$(H_\th,H_\th) = 2,$$
where $\th$ is the maximal root of $\g$ and
$H_\th$ is the element of $\h$ defined by
$$\th(H)=(H_\th,H)\hbox{ for all }H\in\h.$$
%
The affine Lie algebra $\gh$ associated with $\g$ is defined by
$$
\gh = \g\otimes\C((\t))\oplus\C c,
$$
where $c$ is an central element of $\gh$
and the Lie algebra structure is given by
$$
[X\otimes f(\t),Y\otimes g(\t)] = [X,Y]\otimes f(t)g(t) +
c\cdot (X,Y)\, \mathop{\Res}_{\t = 0} (g(t)\cdot df(t)),
$$
for
$X,Y \in \g,\  f(\t),g(\t) \in \Cx.$
We define the subalgebras $\gh_+,\gh_-$ of $\gh$ by
$$
\gh_+=\g\otimes\C[[t]]t,\ \gh_-=\g\otimes\C[t^{-1}]t^{-1}.
$$
Let $P_+$ be the set of all dominant integral weights of $\g$
and for a fixed positive integer $\ell$ (called the level) put
$$
P_\l = \{\ \lm\in P_+\ |\  0\le \lm(H_\th) \le \ell\ \}.
$$
\proclaim
{Proposition 1.1}
For each $\lm \in P_\ell$ there exists a unique irreducible left
$\gh$-module\ $\H _{\lm}$
 (called the integrable highest weight $\gh$-module ) satisfying
 the following properties:

(1)
$V_{\lm} :=  \{\ |v\> \in \H_{\lm}\ |\
\gh_+|v\>= 0 \;\}$
is the irreducible left $\g$-module  with highest weight $\lm$.

(2)
The central element c acts on $\H_{\lm}\ as\ \l\cdot id$.

(3)
$\H_{\lm}$ is generated by $V_\lm$ over $\gh_-$ with only one relation
$$
(X_{\th}\otimes \t^{-1})^{\l-(\th,\lm)+1} |\lm\> = 0,
$$
where $|\lm\>$ is the highest weight vector.
\endproclaim
Similarly we define the integrable highest weight right $\gh$-module
$\H_{\lm}^\dagger$.
There is a perfect bilinear pairing
$$
\<\ |\ \>:\Hld\times\Hl\ \to\ \C,
$$
which is $\gh$-invariant:
$$
\<u|av\>=\<ua|v\>\hbox{ for all } \<u|\ \in\ \Hld\ ,
\ |v\>\ \in\ \Hl\hbox{ and }a\ \in\ \gh.
$$

\subhead
\S2. Pointed Stable Curves
\endsubhead
\medskip
\noindent
{\bf Definition 2.1.}
A set of data $\X=(C;Q_1,Q_2,\ldots,Q_N)$ is called an $N$-pointed stable curve
 of genus $g$, if the following conditions are satisfied:

(1) $C$ is a semi-stable curve of genus $g$.

(2) $Q_1,Q_2,\ldots,Q_N$ are non-singular points of the curve $C$.

(3) The $N$-pointed curve $\X$ has no infinitesimal automorphisms.

\medskip
In the following we also assume the following condition $(*)$\, :
\medskip
$(*)$ Each irreducible component of $C$ contains at least one $Q_j$.
\medskip
\noindent
For a curve $C$ and a non-singular point $Q$ on $C$,
an isomorphism
$$
\s\ :\ \Oh_{C,Q}\ \mathop{\to}^\sim \C [[\t]]
$$
is called a formal neighborhood of $C$ at $Q$.
Here $\Oh_{C,Q}$ is the ring of formal power series at $Q$.
\medskip
\noindent
{\bf Definition 2.2.}
A set of data $\X=(C;Q_1,\ldots,Q_N;\s_1,\ldots,\s_N)$ is called an
$N$-pointed stable curve of genus $g$ with formal neighborhoods if
the following conditions are satisfied:

(1) $(C;Q_1,\ldots,Q_N)$ is an $N$-pointed stable curve of genus $g$.

(2) $\s_j$ is an formal neighborhood at $Q_j$.
\medskip
Introducing a parameter $\xi_j$ for each $j$,
we regard $\s_j$ as an isomorphism
$$
\s_j\ :\ \Oh_{C,Q_j}\ \mathop{\to}^\sim\ \C [[\xi_j]].
$$

\proclaim
{Lemma 2.3}
Let $\X=(C;Q_1,\ldots,Q_N;\s_1,\ldots,\s_N)$ be an N-pointed stable curve
 of genus $g$ with formal neighborhoods which satisfies the condition $(*)$.
Then $\s_1\ldots, \s_N$ induce the following injective homomorphism:
$$
s\ 
:\ H^0(C,\O_C (*\sum_{j=1}^{N}Q_j))\
\to\ \bigoplus_{j=1}^{N}\Cj\, .
$$
\endproclaim
\par
\subhead
\S3. The space of conformal blocks attached to $\X$
\endsubhead
\medskip
We define a Lie algebra
$$
 \gh_N= \oplus_{j=1}^{N}\left(\g\otimes \Cj \right)\oplus \C c
$$
with the following commutation relations:
$$\align
[& \oplus_{j=1}^N X_j\otimes f_j,\oplus_{j=1}^N Y_j\otimes g_j]\ = \\
& \oplus_{j=1}^N [X_j,Y_j]\otimes f_jg_j\ +\ \sum_{j=1}^N (X_j,Y_j)
\mathop{\Res}_{\xi_j=0} (g_j\cdot df_j)\cdot c, \\
&\ \ \ \ \ \ \ c\in \hbox{ center }.
\endalign
$$
\medskip
We also put
$$
\gh(\X)\ =\ \g\otimes_{\C} H^0(C,\O_C (*\sum_{j=1}^{N}Q_j))
$$
and regard it as a subspace of $\gh_N$ by the mapping $\s$
given in Lemma 2.3.

Then by the residue theorem we have the following lemma.

\proclaim
{Lemma 3.1}
The space $\gh (\X)$ is a Lie subalgebra of $\gh_N$.
\endproclaim
\medskip
For each $\vec\lm= (\lm_1,\ldots,\lm_N)\ \in\ (P_\l)^N$ a left
$\gh_N$-module $\Hv$  are defined by
$$
\Hv\  =\ \H_{\lm_1}\otimes\cdots\otimes\H_{\lm_N}\, .
$$
Similarly a right $\gh_N$-module $\H_\vlm^\dagger$ is defied by
$$
\Hv^\dagger\  =\ \H_{\lm_1}^\dagger
\hat{\otimes}\cdots\hat{\otimes}\H_{\lm_N}^\dagger\, .
$$
We use the notation
$$
|u_1\otimes\cdots\otimes u_N\>=|u_1\>\otimes\cdots\otimes|u_N\>
$$
for $|u_j\>\in\H_{\lm_j}\, (j=1,\ldots,N)$.
The $\gh_N$-action on $\Hv$ is given by
$$
(\oplus_{j=1}^N X_j\otimes f_j)|v_1\otimes\cdots\otimes v_N\>\  =\
\sum_{j=1}^N |v_1\otimes\cdots \otimes v_{j-1}\otimes (X_j\otimes f_j)v_j
\otimes
v_{j+1}\otimes\cdots\otimes v_N\>.
$$
The right action on $\Hv^\dagger$ is defined similarly.
There is a $\gh_N$-invariant perfect bilinear pairing
$$
\<\ |\ \>\ :\ \Hv^\dagger\times \Hv\ \to \ \C.
$$
\medskip
\noindent
{\bf Definition 3.2.}
Put
$$
\align
\Vac_{\vec \lm}(\X)\ & =\ \Hv\ /\gh (\X) \Hv ,\\
\Vac_{\vec \lm}^\dagger (\X)\ & =\ \{\ \<\Psi|\in\Hv^\dagger\ ;\
 \<\Psi|\,\gh(\X)=0\ \}\\
&\cong \Hom_\C(\Vac_{\vlm}(\X),\C)\, .
\endalign
$$
We call $\Vac_{\vec \lm}^\dagger(\X)$ the space of conformal blocks (or the
spa\
ce of vacua)
attached to $\X$.
\medskip
\proclaim
{Theorem 3.3}
  The spaces $\Vac_{\vec \lm}(\X)$ and $ \Vac_{\vec \lm}^\dagger (\X)$ are
 finite dimensional vector spaces.
\endproclaim

The above theorem is fundamental to
the formulation of the WZNW-models over algebraic curves
and proved in [TUY].
We will give an alternative proof of Theorem 3.3 in \S 5.
\subhead
\S4. Main idea
\endsubhead
\medskip
The main idea in our proof of Theorem 3.3 is to substitute the equality (4.1)
 for Gabber's theorem.
We explain this point in the following without details.
For simplicity, we consider the $1$-point case, and
let $\X$ be a $1$-pointed stable curve
and $\lm$ be a weight.
We want to show the finite-dimensionality of the space
$$\Vac_\lm(\X)=\H_\lm/\gh(\X)\H_\lm.$$
First, we introduce the filtration $\{\F_\bullet\}$ on $\H_\lm$ by
$$\align
&\F_m\H_\lm=\{ 0\}\ \ (m<0),\ \F_0\H_\lm=V_\lm,  \\
&\F_m\H_\lm =\F_{m-1}\H_\lm+U(\gh)\F_{m-1}\H_\lm\ \ (m>0),
\endalign
$$
and introduce the induced filtration on $\Vac_\lm(\X)$.

Then what we must show is the following:
\medskip
(i) $\dim_\C \F_m \Vac(\X)<\infty $  for any $m\in\Z$.

(ii) $\dim_\C \F_m \Vac(\X)=\dim_\C\F_{m+1} \Vac(\X)$
for sufficiently large $m\in\Z$.
\medskip
As we shall see in the next section,
(i) follows from  the Riemann-Roch theorem
and the proof of (ii), which is more crucial, is reduced to the following:
\medskip
(ii)'  For each $X\in\g$ and $n\in\Z$, there exists an integer $k$ such that
\medskip
 $\ (X\otimes \xi^n)^k|u\>\equiv 0\
\hbox{ mod } \gh(\X)\H_\lm+ \F_{m-1}\H_\lm\
$
for any $m\in\Z$ and $|u\>\in \F_{m-k}\H_\lm$.
\medskip
For a root vector $X_\al\, (\al\in\triangle)$, we can easily show (ii)'
by the nilpotency of $X_\al\otimes\xi^n$ on $V_\lm$,
but for a Cartan element $H_\al=[X_\al,X_{-\al}]$ we need some trick.
We put $H=H_\al,E=X_\al,F=X_{-\al}$.
Then a simple calculation implies the following equality
$$
\split
& (H\otimes\xi^{n})^s (F\otimes\xi^{n})^t |u\>\ \equiv \\
&\frac{1}{t+1} \{
2(s-1) (H\otimes\xi^{n})^{s-2} (F\otimes\xi^{n})^{t+1}
 E\otimes\xi^{n}
|u\>\\
 &+(H\otimes\xi^{n})^{s-1} (F\otimes\xi^{n})^{t+1} E|u\>
 \}\ \ \hbox{ mod }\gh(\X)\H_\lm+\F_{m-1}\H_\lm
\endsplit \tag 4.1
$$
for any $m\in\Z$ and $|u\>\in \F_{m-s-t}\H_\lm$, where
$s$ and $t$ are integers such that $s\geq1,t\geq0$.
By this formula we can increase
the number $t$ by decreasing the number $s$,
and hence
(ii)' for $X=H$ follows from the nilpotency of $F\otimes \xi^n$.
\subhead
\S5. Proof of Theorem 3.3
\endsubhead
\medskip
We consider the following Lie subalgebras
$$
\align
\gh_N^M\ &=\ \oplus_{j=1}^{N} \g\otimes\C[\xi_j^{-1}]\xi_j^{-M}
 \ \ M=0,1,\ldots,\\
\g[1]\ &=\   \{\ \oplus_{j=1}^N X\otimes \xi_j^0 \ ;\ X\in\g\ \}
\endalign
$$
of $\gh_N$. Note that for a positive integer $M$,
the direct sum of  $\gh_N^M$ and $\g[1]$ is again a Lie subalgebra of $\gh_N$.
Put
$$
\W_M\ =\ \H_\vlm/(\gh_N^M\oplus\g[1])\H_\vlm.
$$
We first prove the following lemma.
\proclaim
{Lemma 5.1}
There exists a positive integer $M$ such that
the finite-dimensionality of $\W_M$ implies that of $\Vac_\vlm(\X)$.
\endproclaim
{\it Proof. }
We introduce a filtration $\{\fF_\bullet\}$ on $\gh_N$ as follows.
$$
\fF_p \gh_N \ =\
\cases
&\oplus_{j=1}^{N}\g\otimes\C[[\xi_j]]\xi_j^{-p}
\oplus \C\cdot c \ \hbox{  for } p\geq 0 \, ,\\
&\oplus_{j=1}^{N}\g\otimes\C[[\xi_j]]\xi_j^{-p}\ \ \ \ \ \ \ \ \ \
\hbox{  for } p<0\, .
\endcases
$$
Then $\Hv$ have the natural filtration induced from that of
$\gh_N$:
$$
\fF_p \Hv \ =\ \fF_pU(\gh_N)\cdot V_\vlm\, ,
$$
where 
$$\align
\fF_p U(\gh_N) \ &=\ \sum_{ p_1+p_2+\cdots+p_i \le p} \fF_{p_1}\gh_N
\fF_{p_2} \gh_N\cdots
\fF_{p_i} \gh_N,\\
V_\vlm\ &=\ V_{\lm_1}\otimes\cdots\otimes V_{\lm_N}.
\endalign
$$
We have
$$\gather
\fF_pU(\gh_N)\cdot \fF_q\Hv\ \subset\ \fF_{p+q}\Hv,
\\
\fF_p \Hv = \{0\} \hbox{ for } p<0.
\endgather
$$
We introduce the induced filtrations on subalgebras of $\gh_N$ and
the quotient filtrations on $\Vac_\vlm(\X)$ and $\W_M$,
and  consider the associated graded objects $\Gr_\bullet (\ )$.
Then we have the following exact sequences of graded vector spaces over $\C$:
$$\align
&0 \to\,\ \ \ \ \ \Gr_\bullet(\gh(\X)\Hv)\ \ \ \ \ \,    \to\ \Gr_\bullet\Hv\
\to    \Gr_\bullet\Vac_\vlm(\X) \ \to\ 0,\\
&0 \to\     \Gr_\bullet((\gh_N^M\oplus \g [1]) \Hv)\ \to\ \Gr_\bullet\Hv\
\to\ \,\ \Gr_\bullet\W_M        \ \to\ 0.
\endalign
$$
The Riemann-Roch theorem implies that,
 for a sufficiently ladrge integer $M$, we have
$$
\Gr_\bullet \left( (\gh_N^M\oplus \g[1])\Hv\right)
\ \subset\ \Gr_\bullet (\gh(\X)\Hv)\tag 5.1
$$
as subspaces of $\Gr_\bullet \Hv$.
Therefore for such an integer $M$,
we have the following surjective homomorphism:
$$
\Gr_\bullet \W_M\ \to\ \Gr_\bullet\Vac_\vlm(\X)\ \to\ 0.
$$
This proves Lemma 5.1.\hfill $\square$
\medskip
In the following we fix an integer $M$ which satisfies (5.1) and put
$\W=\W_M$.
The rest of this section is devoted to prove the finite-dimensionality of
 $\W$.

We define the finite dimensional vector space $\frak{b}$ by
$$
\frak{b}\ =\ \oplus_{j=1}^N \left(
\g\otimes \left(\oplus_{n=0}^{M-1}\C \xi_j^{-n}
\right)\right).
$$
Since  $\gh_N^M$ is an ideal of $\gh_N^0$, a Lie algebra structure on
$\frak{b}$ is defined through the isomorphism
$$
\frak{b}\ \cong\ \gh_N^0/\gh_N^M.
$$
Put
$$
\U \ =\ \Hv / \gh_N^M \Hv.
$$
Then $\U $ has a structure of $U(\frak{b})$-module and it is generated by the
 image
$\tV$ of $V_\vlm$ on $\U$:
$$
\U\ \cong\ U(\frak{b})\cdot\tV.
\tag$5.2$
$$
\remark
{Remark }
In [TUY] the finite-dimensionality of $\U$, which implies that of $W$,
 is proved  by Gabber's theorem.
\endremark
\medskip
Now we define other filtrations $\{\F_{\bullet}\}$ on $U(\gh_N^0)$
and $\Hv$ as follows.
$$\align
\F_mU(\gh_N^0)\ &=\
\cases
\{0\}    \ \ \ \ \ \ \ &m<0,\\
\C\cdot 1\ \ \ \ \ \ \ &m=0,\\
\F_{m-1}U(\gh_N^0)+\gh_N^0\F_{m-1}U(\gh_N^0)\ \ &m> 0,
\endcases \\
\F_m\Hv\ &=\ \F_mU(\gh_N^0)\cdot V_{\vec \lm}.
\endalign
$$
Then we have
$$
\align
[\F_mU(\gh_N^0),\F_nU(\gh_N^0)]&\subset \F_{m+n-1}U(\gh_N^0),\\
\F_mU(\gh_N^0)\F_n\Hv&\subset \F_{m+n}\Hv.
\endalign
$$
On subspaces of $\Hv$
we introduce the induced filtrations from that of $\Hv$.
On $\Vac_{\vec \lm}(\X)$, $\W$ and $\U$ we introduce
the quotient filtrations from that of $\Hv$.
On $U(\frak{b})\cong U(\gh_N^0)/\gh_N^MU(\gh_N^0)$ we introduce the quotient
 filtrations from that of $U(\gh_N^0)$.

We denote the associated graded objects by $\gr_\bullet(\ )$.
Then we have the following isomorphism as graded algebras:
$$
\gr_\bullet U(\frak{b})\  \cong\ S(\frak{b}),
$$
where $S(\frak{b})$ is the symmetric algebra of $\frak{b}$.
By (5.2), the space $\gr_\bullet\U$ is
generated by $\gr_0\U$ as a $S(\frak{b})$-module:
$$
\gr_\bullet \U\ \cong\ S(\frak{b})\cdot \gr_0 \U.
$$
We denote the degree $m$-part of $S(\frak{b})$ by
$S^m(\frak{b})$.
Then we have the following lemma.
\proclaim
{Lemma 5.2}
For each integer $m$, there exists a commutative diagram of surjective
homomorphisms:
$$\CD
\F_m \U\ @>{\rho_m}>> S^m(\frak{b})\cdot \gr_0 \U \\
@VVV          @V{\varphi_m}VV\\
\F_m \W  @>>> \gr_m \W\, .
\endCD\
$$
\hfill $\square$
\endproclaim
This lemma implies that the space
$\gr_m\W=\varphi_m\left(S^m(\frak{b})\cdot \gr_0\U\right)$
 is finite dimensional over $\C$,
since $\frak{b}$ and $\gr_0\U$ are finite dimensional.
Hence, to prove the finite-dimensionality of  $\W$,
it is sufficient to prove the following.
\proclaim
{Claim 1}
For a sufficiently large integer $m$, we have
$$
\gr_m\W=\{0\}.
$$
\endproclaim
Put
$$\pi_m=\varphi_m\circ\rho_m \ :\ \F_m\U\ \to \ \gr_m\W.
$$

To prove Claim 1, it is sufficient to prove the following.
\proclaim
{Claim 2}
For each $X\in \g$, there exists an integer $K$ such that we have
$$
k\geq K \Rightarrow \pi_m((X\otimes\xi_j^{-n})^k  |u\>)\ =\ 0\tag{5.3}
$$
for any  $m\in\Z,\;
j=1,\ldots,N,\; n=0,1,\ldots,M-1 $ and $|u\>\in \F_{m-k} \U$.
\endproclaim
Before proving Claim 2, let us show that Claim 1 follows from  Claim 2.
Assuming Claim 2,
we can take an integer $K$ for which (5.3) holds for any
 $X\in\g,m\in\Z,\; j=0,\ldots,N,\; n=0,\ldots,M-1$ and $|u\>\in\F_{m-k}\U$.
Fix an integer $m$ larger than $K\times\dim \frak{b}=
K\times\dim\g\times N\times M$,
and take an element
$$
a \ =\ \prod_{i=1}^{\dim\g}\prod_{j=1}^{N}\prod_{n=0}^{M-1}
(J^i\otimes\xi_j^{-n})^{k_{i,j,n}}
$$
of $S^m(\frak{b})$ with  $\sum_{i,j,n} k_{i,j,n}=m$,
where
$\{J^i;i=1,\ldots,\dim \g\}$ is a basis of $\g$.
Then we can find at least one index $k_{i',j',n'}$
which is larger than $K$,
and by the assumption we have  $\varphi_m (a |v\>)=0$ for any $|v\> \in
\gr_0\U\
$.
Therefore we have $ \gr_m\W =\varphi_m(S^m(\frak{b})\cdot\gr_0\U)=\{0\}$.

%
%

Let us prove Claim 2.
Fix integers $j=1,\ldots,N ,n=0,\ldots,M-1$ and a positive root
$\alpha\in \triangle_+$, and put
$$
E=X_\al,\ F=X_{-\al},\ H=H_\al=[X_\al,X_{-\al}]\, .
$$
For $X=E$ or $F$, it is easy to show Claim 2,
since $\Xj^{-n}$ acts nilpotently on $V_\vlm$.
In order to prove Claim 2 for $X=H$,
we consider the element $|u\>$ in $\F_m\U$ of the form
$$
|u\> = \ (H\j^{-n})^{s-1} (F\j^{-n})^{t+1} |v\>,
$$
where $s$ and $t$ are integers such that $s\geq 1$ and $t\geq 0$, and
$ |v\>\in \F_{m-s-t}\U$.

Put
$$E[1]=\oplus_{j=1}^N E\otimes\xi_j^0.
$$
Then we have
$$
\align
&E[1] |u\>\ =\ \oplus_{i=1}^N E\otimes\xi_i^0 |u\>\ =\ \tag 5.4 \\
 [E\otimes\xi_j^0, (H&\j^{-n})^{s-1} (F\j^{-n})^{t+1}] |v\>
+ (H\j^{-n})^{s-1} (F\j^{-n})^{t+1}E[1] |v\> .
\endalign
$$
Note that
$$E[1]|v\>\in \F_{m-s-t}\U.
$$
Sending (5.4) by $\pi_m$ after calculating the commutator, we have
$$
\align
\pi_m(E[1]|u\>)\ =\ &-2(s-1) \pi_m \left( (H\j^{-n})^{s-2}
(F\j^{-n})^{t+1} E\j^{-n} |v\>\right)\\
&+ (t+1) \pi_m \left( (H\j^{-n})^s (F\j^{-n})^t |v\> \right)\\
&+ \pi_m \left( (H\j^{-n})^{s-1} (F\j^{-n})^{t+1}E[1] |v\> \right).
\endalign
$$
On the other hand, by the definition of $\W$,
we have $\pi_m(E[1]|u\>)=0$.
Therefore we get the following formula:
$$
\align
\pi_m& \left( (H\j^{-n})^s (F\j^{-n})^t |v\> \right)\ =\\
&\frac{1}{t+1} \{
2(s-1) \pi_m \left( (H\j^{-n})^{s-2} (F\j^{-n})^{t+1} E\j^{-n}
|v\>\right)\\
 &+ \pi_m\left((H\j^{-n})^{s-1} (F\j^{-n})^{t+1} E[1]|v\> \right)
 \}.
\endalign
$$

By this formula we can increase the number $t$ by decreasing the number $s$.
 Hence if we take an integer $K'$ such that
$$
 k\geq K'\Rightarrow \pi_m\left((F\j^{-n})^{k}|v\>\right)\
=\ 0 \hbox{ for any }
m\in\Z \hbox{ and } |v\>\in \F_{m-k}\U,
$$
then we have
$$
 k\geq 2K'\Rightarrow \pi_m\left((H\j^{-n})^{k}|v\>\right)\
=\ 0 \hbox{ for any }
m\in\Z \hbox{ and } |v\>\in \F_{m-k}\U.
$$
This proves Claim 2 for $X=H$. Now, it is easy to prove Claim 2 for any $X$.
\hfill $\square$

\par
\medskip
{\bf Acknowledgements}
\medskip
I would like to thank M. Kashiwara and  T. Miwa
for a careful reading of the manuscript and helpful advice.
I am also grateful to A. Tsuchiya for important suggestions,
and to K. Ueno and H. Ooguri for useful comments.


\Refs
\widestnumber\key{W-W}

\ref
\key Ga \by O. Gabber
\paper The integrability of the characteristic variety
\jour Amer. J. Math.\vol 103 \yr 1981 \pages 445
\endref


\ref
\key Ka \by V. Kac
\book Infinite dimensional Lie algebras
\publ Cambridge University Press., Third edition \yr 1990
\endref


\ref \key TK \by A. Tsuchiya and Y. Kanie
\paper Vertex operators in conformal field theory on $\P^1$ and monodromy
representations of braid group
\jour Adv. Stud. in Pure Math.\vol 16 \yr 1988 \pages 297
\endref


\ref
\key TUY \by A. Tsuchiya, K. Ueno and Y. Yamada
\paper Conformal field theory on universal family of stable curves with gauge
symmetries
\jour Adv. Stud. in Pure Math. \vol 19 \yr 1989 \pages 459--566
\endref


\endRefs
\enddocument
\end